\documentclass[10pt]{article}
\usepackage{jheppub}
\allowdisplaybreaks[1]

\newcommand{\ba}{\begin{eqnarray}}
\newcommand{\ea}{\end{eqnarray}}

\newcommand{\cf}{c_f}
\newcommand{\nf}{n_f}
\newcommand{\str}{{\rm sTr}}
\newcommand{\la}[1]{\label{#1}}

\newcommand{\eq}{eq.~}
\newcommand{\se}{section~}
\newcommand{\secs}{sections~}
\newcommand{\app}{appendix~}
\newcommand{\eqs}{eqs.~}
\newcommand{\nr}[1]{(\ref{#1})}
\newcommand{\ep}{\varepsilon}
\newcommand{\CA}{C_{\mathrm{A}}}
\newcommand{\CF}{C_{\mathrm{F}}}
\newcommand{\NA}{N_{\mathrm{A}}}
\newcommand{\NF}{N_{\mathrm{F}}}
\newcommand{\TF}{T_{\mathrm{F}}}
\newcommand{\Nf}{N_{\mathrm{f}}}
\newcommand{\order}[1]{{\cal O}(#1)}

\newcommand{\code}[1]{{\tt #1}}

\newcommand{\gammaGG}{\gamma_3}
\newcommand{\gammaGGG}{\gamma_1^{3g}}
\newcommand{\gammaGGGG}{\gamma_1^{4g}}
\newcommand{\gammaCC}{\gamma_3^{c}}
\newcommand{\gammaCc}[1]{\gamma_{3#1}^{c}}
\newcommand{\gammaCCG}{\gamma_1^{ccg}}
\newcommand{\gammaCcg}[1]{\gamma_{1#1}^{ccg}}
\newcommand{\gammaQQ}{\gamma_2}
\newcommand{\gammaQq}[1]{\gamma_{2#1}}
\newcommand{\gammaQQG}{\gamma_1^{\psi\psi g}}
\newcommand{\MSbar}{${\overline{\mbox{\rm{MS}}}}$}
\newcommand{\slashed}[1]{#1\!\!\!/}
\newcommand{\m}{M}

\title{The five-loop Beta function for a general gauge group and anomalous dimensions beyond Feynman gauge}

\preprint{\mbox{}\hfill BI-TP 2017/13\\\mbox{}\hfill DESY 17-142\\\mbox{}\hfill IPPP/17/68}

\author[a]{Thomas Luthe,}
\author[b]{Andreas Maier,}
\author[c]{Peter Marquard}
\author[d]{and York Schr\"oder}

\affiliation[a]{Faculty of Physics, University of Bielefeld, 33501 Bielefeld, Germany}
\affiliation[b]{Institute for Particle Physics Phenomenology, Durham University, Durham, United Kingdom}
\affiliation[c]{Deutsches Elektronen Synchrotron (DESY), Platanenallee 6, Zeuthen, Germany}
\affiliation[d]{Grupo de Cosmolog\'ia y Part\'iculas Elementales, Universidad del B\'io-B\'io, Casilla 447, Chill\'an, Chile}

\emailAdd{tluthe@physik.uni-bielefeld.de}
\emailAdd{andreas.maier@durham.ac.uk}
\emailAdd{peter.marquard@desy.de}
\emailAdd{yschroder@ubiobio.cl}

\abstract{We focus on a non-abelian gauge field coupled to a single (but general) representation of a family of $\Nf$ fermions. By using the same machinery that had allowed us to evaluate the sub-leading large-$\Nf$ term of the five-loop Beta function earlier, we here report on a confirmation of the all-$\Nf$ result that has in the meantime been published by another group. Furthermore, in order to push forward the 5-loop renormalization program regarding gauge parameter dependence, we present the linear terms of the complete set of anomalous dimensions, in an expansion in the covariant gauge parameter around the Feynman gauge.}

\begin{document}
\maketitle

%%%%%%%%%%%%%%%%%%%%%%%% SECTION %%%%%%%%%%%%%%%%%%%%%%
%
\section{Introduction}
\la{se:intro}

In modern high-energy physics experiments, in order to closely scrutinize (and eventually go beyond) our established particle physics models such as the Standard Model (SM), it is important to push the precision of theoretical predictions that follow from these models to the highest possible level. All parameters that appear in these quantum field theories such as the SM change as functions of the energy scale, in a well-defined way that is governed by so-called renormalization group equations. These, in turn, depend on a number of renormalization group parameters that can be deduced from the underlying quantum field theory. 

Perhaps the most fundamental of such renormalization group parameters is the Beta function, governing the running of the gauge coupling constant, and consequently much effort has been invested into precision determinations of this coefficient. After seminal work at one-loop order \cite{Gross:1973id,Politzer:1973fx}, demonstrating the asymptotically free nature of the strong coupling constant and therefore establishing Quantum Chromodynamics (QCD) as a central part of the Standard Model, perturbative corrections have been pushed to 2-loop \cite{Caswell:1974gg,Jones:1974mm}, 3-loop \cite{Tarasov:1980au,Larin:1993tp} and 4-loop \cite{vanRitbergen:1997va,Czakon:2004bu} level. Five-loop results have appeared over the last ten years or so, first for the case of Quantum Electrodynamics (QED) \cite{Baikov:2008cp,Baikov:2010je,Baikov:2012zm}, then for physical QCD with gauge group SU(3) \cite{Baikov:2016tgj,Chetyrkin:2016uhw}, and finally for QCD with more general gauge groups \cite{Luthe:2016ima,Herzog:2017ohr}.

Given the complexity of the five-loop calculation, there is an urgent need to confirm the Beta function as given in \cite{Herzog:2017ohr} by an independent approach. We fill this gap in the present paper, building upon our earlier work \cite{Luthe:2016ima}, where a proof-of-concept had been laid out (and in the meantime been successfully tested and expanded, see \cite{Luthe:2016xec,Luthe:2017ttc}). 
Throughout the paper, we work in dimensional regularization around $d=4-2\ep$ space-time dimensions and in the \MSbar{} scheme.

Of course, the (gauge-invariant) Beta function is not the only fundamental parameter governing renormalization of a gauge theory. All fields and parameters of the theory need to be renormalized, giving rise to a set of renormalization constants (RCs) that can be evaluated order by order in perturbation theory. 
Perhaps the second most important representative of this set is the (gauge-invariant) renormalization constant for the quark mass, needed for a precise evolution of measured low-energy quark masses to current and future high-energy collider experiment energies. It has been known at two \cite{Tarrach:1980up} and three loops \cite{Tarasov:1982gk,Larin:1993tq} for a long time already; at four loops, complete results for SU($N$) and QED as well as general Lie groups are available \cite{Chetyrkin:1997dh,Vermaseren:1997fq}; at five loops, mass renormalization is known for SU(3) as well as general Lie groups \cite{Baikov:2014qja,Luthe:2016xec,Baikov:2017ujl}.

The remaining members of the set of RCs depend on the gauge parameter. 
At four loops, these are known since more than a decade for SU($N$) and Lie groups, see \cite{Chetyrkin:2004mf,Czakon:2004bu} and references therein. 
Full gauge dependence for the case of Lie groups has been added only recently \cite{Luthe:2016xec,Luthe:2017ttc}.
At five loops and for a general Lie group, all of them are presently known in Feynman gauge \cite{Luthe:2016xec,Luthe:2017ttc} (some notable exceptions being all-order Landau gauge results in the limit of many fermion flavors, see e.g.\ \cite{Gracey:1993ua}).
In order to push forward the renormalization program, we continue to evaluate corrections to the set of renormalization constants, which at five-loop level are available (mostly) in Feynman gauge only. To this end, we present new results for the complete set of RCs, including linear terms in the gauge parameter, in an expansion around the Feynman gauge. These types of terms might be needed in future projects, to provide for valuable cross-checks concerning gauge invariance of the observables under investigation. 

The structure of the paper is as follows. We begin by explaining our calculational setup in \se\ref{se:setup}. There, we define the set of renormalization constants and anomalous dimensions we are after, describe the massive regularization method we have employed to extract ultraviolet divergences, and introduce the set of group invariants that are needed to express the higher-order results. In \se\ref{se:beta}, we present our results for the five-loop gluon field anomalous dimension (in Feynman gauge), from which we extract the (gauge-invariant) Beta function. We then continue to push the 5-loop renormalization program further, and present the linear terms in an expansion around the Feynman gauge in \se\ref{se:beyond}, and conclude in \se\ref{se:conclu}. Two appendices are devoted to list perturbative coefficients for certain renormalization constants that are needed in the main text.

%%%%%%%%%%%%%%%%%%%%%%%% SECTION %%%%%%%%%%%%%%%%%%%%%%
%
\section{Setup}
\la{se:setup}

We begin by making a number of definitions and technical remarks. First, we define the various renormalization constants and anomalous dimensions that are the focus of this work, and list relations between them. Then, we explain parts of our computational setup that allows us to extract these coefficients from the short-distance (ultraviolet) divergences of the theory. Finally, we introduce some convenient definitions for gauge group invariants that will allow us to compactly present results later on.

%%%%%%%%%%%%%%%%%%%%%%%% SUBSECTION %%%%%%%%%%%%%%%%%%%%%%
%
\subsection{Renormalization constants}

The fermion-, gauge- and ghost fields as well as fermion mass, gauge coupling and gauge-fixing parameter of the gauge theory are renormalized multiplicatively via 
\ba
\psi_b &=& \sqrt{Z_2}\psi_r \;,\quad A_b = \sqrt{Z_3}A_r \;,\quad c_b = \sqrt{Z_3^c}c_r \;,\\
\la{eq:ZgDef}
m_b &=& Z_m m_r \;,\quad g_b = \mu^\ep Z_g g_r \;,\quad \xi_{L,b} = Z_\xi \xi_{L,r} \;.
\ea
We have used the subscript $b$ and $r$ for bare  and renormalized quantities, respectively. All renormalization constants (RCs) have the form $Z_i=1+{\cal O}(g_r^2)$. There actually is no need to renormalize the gauge-fixing term $\sim(\partial A)^2/\xi_L$, such that setting $Z_\xi=Z_3$ leaves us with five independent renormalization constants only. 
A very economic way of recording the various renormalization constants $Z_i$ is to merely list the corresponding anomalous dimensions, defined by 
\ba
\la{eq:gammaDef}
\gamma_i &=& -\partial_{\ln\mu^2}\ln Z_i \;.
\ea

Following usual conventions, instead of considering $Z_g$, one renormalizes the gauge coupling squared (which in our notation is 
\ba
\la{eq:a}
a\equiv\frac{\CA\,g_r^2(\mu)}{16\pi^2}
\ea
with $\CA$ the quadratic Casimir operator of the adjoint representation of the gauge group, cf.\ \se\ref{se:color}) with the factor $Z_a\equiv Z_g^{\,2}$ and calls the corresponding anomalous dimension $\gamma_a=2\gamma_g\equiv\beta$ the Beta function. Note that, due to the renormalization scale independence of the bare gauge coupling, using \eqs\nr{eq:ZgDef} and \nr{eq:gammaDef} this immediately implies
\ba
\la{eq:betaDef}
\beta=\ep+\partial_{\ln\mu^2}\ln a \quad\Leftrightarrow\quad \partial_{\ln\mu^2}a = -a\Big[\ep-\beta\Big]\;.
\ea
The Beta function is a gauge invariant object and is known at five loops \cite{Herzog:2017ohr}, as discussed further in \se\ref{se:beta}. The second gauge invariant anomalous dimension is $\gamma_m$, corresponding to the renormalization of the quark mass. At the five-loop level, it has been given in \cite{Luthe:2016xec}, and confirmed by \cite{Baikov:2017ujl}.

To complete the renormalization program, we are left with choosing (besides the gauge invariants $\beta$ and $\gamma_m$) three further RCs. These three coefficients will necessarily be gauge-parameter dependent, and at the five-loop level only the Feynman-gauge results are known so far, see \cite{Luthe:2017ttc} for a complete list of results.
In practical calculations, it can sometimes be convenient to consider 'vertex RCs' which are products of the $Z_i$, such as those that multiply the 3-gluon, 4-gluon, ghost-gluon and quark-gluon vertex. These vertex RCs are usually denoted as $Z_1^j$ (where $j\in\{3g,4g,ccg,\psi\psi g\}$). Out of this set, we found it convenient to evaluate the combination $Z_1^{ccg}=\sqrt{Z_3}\,Z_3^c\,Z_g$, giving us the anomalous dimension $\gammaCCG$.
For the remaining two of the minimal set of five RCs, we simply pick $Z_2$ and $Z_3^c$, encoded in the respective anomalous dimensions $\gammaQQ$ and $\gammaCC$.

Once the minimal set of renormalization constants (chosen here to be $\gamma_m$, $\beta$, $\gammaCC$, $\gammaCCG$ and $\gammaQQ$, as explained above) is known, all other anomalous dimensions can be reconstructed from simple linear relations, since they are related via gauge invariance of the QCD action (see e.g.\ \cite{Chetyrkin:2004mf}): 
\ba
\la{eq:gammaGG}
\gammaGG\;=&2(\gammaCCG-\gammaCC)-\beta \;,\quad
\gammaGGG&=\;3(\gammaCCG-\gammaCC)-\beta \;,\\
\la{eq:gammaGGGG}
\gammaGGGG\;=&4(\gammaCCG-\gammaCC)-\beta \;,\quad
\gammaQQG&=\;\gammaCCG-\gammaCC+\gammaQQ \;.
\ea

If one needs to reconstruct renormalization constants $Z_i$ from the anomalous dimensions $\gamma_i$, one can start from \eq\nr{eq:gammaDef}, recalling that $Z_i(a,\xi_L)$ depends on the renormalization scale through both of its variables. Using the $d$\/-dimensional Beta function of \eq\nr{eq:betaDef}; remembering that the gauge parameter renormalizes as the gluon field $\xi_{L,b}=Z_3\xi_{L,r}$; expressing the gauge parameter as $\xi_L=1-\xi$ where $\xi=0$ now corresponds to Feynman gauge; and converting all anomalous dimensions to our preferred minimal set, one obtains the relation
\ba
\la{eq:g2Z}
\gamma_i &=& 
-a(\beta-\ep)(\partial_a\ln Z_i) -(2\gammaCCG-2\gammaCC-\beta)(\xi-1)(\partial_\xi\ln Z_i) \;.
\ea
The coefficients $z_i^{(n)}$ of the RCs $Z_i=1+\sum_{n>0}z_i^{(n)}/\ep^n$ finally follow from solving \eq\nr{eq:g2Z}, requiring $\gammaCCG$, $\gammaCC$ and $\beta$ at one loop lower only.
In turn, once the RCs $Z_i$ are available, the corresponding anomalous dimensions can be extracted from the single poles, $\gamma_i=a\partial_a z_i^{(1)}$.

%%%%%%%%%%%%%%%%%%%%%%%% SUBSECTION %%%%%%%%%%%%%%%%%%%%%%
%
\subsection{Extraction of ultraviolet divergences}
\la{se:uv}

In order to compute the field, mass, and vertex renormalization constants in
the \MSbar{} scheme we are tasked with extracting the ultraviolet (UV) divergences of
corresponding Green's functions. Since UV divergences are known to be
independent of the masses and external momenta, it is desirable to
eliminate as many of these scales as possible to facilitate the computation. In
fact, in the calculationally most efficient approaches all scales are
initially sent to zero and auxiliary masses are only introduced to
separate infrared (IR) from UV divergences.

One highly successful method for infrared regularization is given by the
R$^*$\/-operation~\cite{Chetyrkin:1982nn,Chetyrkin:1984xa,Chetyrkin:2017ppe,Herzog:2017bjx}. For
instance, it has been used in the recent computations of the five-loop
anomalous dimensions in QCD~\cite{Baikov:2014qja,Baikov:2016tgj} and
their generalization to an arbitrary gauge
group~\cite{Herzog:2017ohr,Baikov:2017ujl}. Its main appeal is that
$L$-loop anomalous dimensions can be deduced from the calculation of
$(L-1)$-loop massless diagrams with one external momentum. The price to
pay is an increased conceptual complexity. Up to now, only the ``local''
variant of the R$^*$\/-operation has been automatized \cite{Herzog:2017bjx}, whereas the
computationally more efficient ``global'' operation still requires
significant manual work.

In this work, we will pursue a conceptionally much simpler alternative
approach pioneered in~\cite{Misiak:1994zw,Chetyrkin:1997fm}, which is
also sufficiently powerful to allow the computation of five-loop
anomalous dimensions in a general gauge
group~\cite{Luthe:2016ima,Luthe:2016xec,Luthe:2017ttc}.\footnote{See
  also~\cite{vanRitbergen:1997va,Vermaseren:1997fq,Czakon:2004bu,Zoller:2015tha,Chetyrkin:2016ruf,Zoller:2016sgq,Chetyrkin:2017mwp,Zerf:2017zqi}
  for applications of this method to four-loop problems.}
It is based on the exact decomposition~\cite{Chetyrkin:1997fm}
\begin{equation}
  \la{eq:Kostja_dec}
  \frac{1}{(l+q)^2} = \frac{1}{l^2 - \m^2} - \frac{q^2 + 2lq + \m^2}{(l^2 - \m^2)(l+q)^2}\,,
\end{equation}
where $l$ is a linear combination of loop momenta, $q$ a linear
combination of external momenta. By introducing an auxiliary mass $\m$ we
have ensured that the first term on the right-hand side is IR
finite. While the second term can lead to IR divergences, the UV degree
of divergence is reduced compared to both other terms.

After subtracting all subdivergences, we can apply
relation~(\ref{eq:Kostja_dec}) iteratively to all massless propagators
in order to decompose a given diagram into a UV-divergent part
containing only massive denominators and a UV-finite but potentially
IR-divergent remainder. Since we are only interested in the UV
divergence, we can safely drop this remainder.

Furthermore, we note that the left-hand side of
Eq.~(\ref{eq:Kostja_dec}) is independent of the auxiliary mass $\m$ and
the dependence on $\m$ therefore has to drop out in the final
result. This allows us to discard the factor $\m^2$ in the numerator of
Eq.~(\ref{eq:Kostja_dec}) and cancel the resulting $\m$-dependence using
new counterterms proportional to $\m^2$. Any such counterterm originating
from a four-dimensional Lagrangian has to be proportional to $\m^2 A^2$
and can therefore be interpreted as a ``gauge boson mass'' counterterm. The
normalization is then fixed by the condition that the inverse gauge boson
propagator must not contain a (UV-divergent) contribution proportional
to $\m^2$.

We note that the prescription of repeatedly applying
Eq.~(\ref{eq:Kostja_dec}) and discarding all terms that are UV finite or
contain factors of $\m^2$ can be formulated in an even simpler way. By
direct comparison, we see that it is completely equivalent to introduce
the auxiliary mass $\m$ in all denominators and perform a Taylor
expansion in small external momenta.

It is then straightforward to modify the denominators and introduce the
gauge boson mass counterterm by changing the Feynman rules for the
propagators accordingly. We start from the conventionally renormalized
propagators ${\cal D}^\delta$, where $\delta=f,c,g$ denotes a fermion,
ghost, or gauge boson, respectively. They are given by the usual expressions (we drop the subscripts  $r$ of \eq\nr{eq:ZgDef} from now on, writing $m$ for the renormalized fermion mass and $\xi_L$ for the renormalized gauge parameter)
\ba
  Z_2 \,{\cal D}^f(p) ={} \frac{i}{\slashed{p} - Z_{m} m} \;,\quad
  Z_3^c \,{\cal D}^c(p) ={} \frac{i}{p^2} \;,\quad
  Z_3 \,{\cal D}^g_{\mu\nu}(p) ={} -\frac{i}{p^2}\bigg[g_{\mu\nu} - (1 - Z_3 \xi_L)
                             \frac{p_\mu p_\nu}{p^2}\bigg]\,.
\ea
In order to eliminate a scale, we expand the fermion propagator in
the limit of a small fermion mass. Retaining the first two terms in
the expansion is sufficient for the determination of the wave function
and mass renormalization constants. Introducing an auxiliary mass $\m$
and the corresponding gauge boson mass counterterm $Z_{\m^2}$ we then
obtain
\ba
  \tilde{\cal D}^f(p) &=&  i \,\frac{\slashed{p} + m}{p^2-\m^2}\Big(1 + i \big[\delta
  Z_{2} (\slashed{p} - m) -  Z_2 \,\delta Z_{m}\, m \big] \tilde{\cal D}^f(p)\Big) +
                           {\cal O}(m^2)\;,\\
  \tilde{\cal D}^c(p) &=& \frac{i}{p^2-\m^2}\Big(1 + i \,\delta Z_3^c\, p^2
                           \tilde{\cal D}^c(p)\Big)\;,\\
  \tilde{\cal D}^g_{\mu\nu}(p) &=& \frac{-i}{p^2-\m^2}\Big(g_{\mu\rho} - \xi\, \frac{p_\mu p_\rho}{p^2
                    - \m^2}\Big)\Big(g_{\rho\nu} - i\big[
                     Z_{\m^2} \m^2 g_{\rho\sigma} + \delta Z_3
                    (g_{\rho\sigma}p^2 - p_\rho p_\sigma)\big]\tilde{\cal D}^g_{\sigma\nu}(p)\Big) 
\ea
for the modified propagators $\tilde{\cal D}$. 
We have again rewritten $\xi_L=1-\xi$, such that $\xi=0$ corresponds to Feynman gauge.
The counterterm $Z_{\m^2}$ is then fixed by requiring that the gauge boson
propagator with vanishing external momentum must be finite at each order
in perturbation theory. For future reference, we list this auxiliary counterterm $Z_{\m^2}$ in \app\ref{app:ZM}.

After the above-mentioned Taylor expansion in small external momenta, we are left with fully massive vacuum integrals, where all propagators share the common regulator mass $\m$. We have two independent in-house codes at hand, \code{crusher} \cite{crusher} and \code{Spades} \cite{Luthe:2015ngq}, that systematically exploit the well-known integration-by-parts identities to achieve the mapping of such fully massive vacuum integrals onto a small set of master integrals.
At five loops, these master integrals have recently been evaluated to high numerical precision \cite{Luthe:2015ngq}, allowing high-confidence analytic fits of individual integrals, sums thereof, and/or full results. For further details on our reduction and integration strategy and all relevant references, we refer to \cite{Luthe:2016ima,Luthe:2016xec,Luthe:2017ttc,Luthe:2016sya,Luthe:2016spi}.

%%%%%%%%%%%%%%%%%%%%%%%% SUBSECTION %%%%%%%%%%%%%%%%%%%%%%
%
\subsection{Notation for color factors}
\la{se:color}

Let us finally define some useful notation concerning group invariants, which we will need to present our results. 
To this end, we re-iterate notation that we had already utilized in previous works \cite{Luthe:2016ima,Luthe:2016xec,Luthe:2017ttc}. 
We focus on a Yang-Mills theory coupled to $\Nf$ fermions in the fundamental representation. It is straightforward to generalize our results to fermions in a (single) arbitrary representation $R$ by substituting all generators of the fundamental representation with generators of $R$.

The real and antisymmetric structure constants $f^{abc}$ are defined by the commutation relations $T^aT^b-T^bT^a=i f^{abc}T^c$ between hermitian generators $T^a$ of a semi-simple Lie algebra, with trace normalization ${\rm Tr}(T^aT^b)=\TF\delta^{ab}$.
The quadratic Casimir operators of the fundamental and adjoint representations (of dimensions $\NF$ and $\NA$, respectively) are then defined in the usual way, as $T^aT^a=\CF1\!\!1$ and $f^{acd}f^{bcd}=\CA\delta^{ab}$.
To facilitate compact representations of our results, we find it convenient to use the following normalized combinations of group invariants:
\ba
\nf=\frac{\Nf\,\TF}{\CA} \quad,\quad 
\cf=\frac{\CF}{\CA} \;.
\ea

In loop diagrams, one typically encounters traces of more than two group generators, giving rise to higher-order group invariants. These higher-order traces can be systematically classified in terms of combinations of symmetric tensors \cite{vanRitbergen:1998pn}. Rewriting the generators of the adjoint representation as $[F^a]_{bc}=-if^{abc}$, we need the following three combinations (again, we normalize conveniently):
\ba
d_1=\frac{[\str(T^aT^bT^cT^d)]^2}{\NA\TF^2\CA^2} \;,\;
d_2=\frac{\str(T^aT^bT^cT^d)\,\str(F^aF^bF^cF^d)}{\NA\TF\CA^3} \;,\;
d_3=\frac{[\str(F^aF^bF^cF^d)]^2}{\NA\CA^4} \,.\;\;
\ea
Here, $\str(ABCD)=\frac16{\rm Tr}(ABCD+ABDC+ACBD+ACDB+ADBC+ADCB)$ is a fully symmetrized trace.

Taking the gauge group to be SU($N$) and setting $\TF=\frac12$ and $\CA=N$, our set of normalized invariants then reads \cite{vanRitbergen:1998pn}
\ba
\la{eq:sun}
\nf=\frac{\Nf}{2N}
\;,\;\;
\cf=\frac{N^2-1}{2N^2}
\;,\;\;
d_1=\frac{N^4-6N^2+18}{24N^4}
\;,\;\;
d_2=\frac{N^2+6}{24N^2}
\;,\;\;
d_3=\frac{N^2+36}{24N^2}\;.\quad
\ea
From here, one can for example easily obtain the SU(3) coefficients, corresponding to physical QCD.

%%%%%%%%%%%%%%%%%%%%%%%% SECTION %%%%%%%%%%%%%%%%%%%%%%
%
\section{Gauge field anomalous dimension and Beta function}
\la{se:beta}

Following up on our previous work on the $\Nf^{\{4,3\}}$ terms of the five-loop Beta function \cite{Luthe:2016ima} as well as our determinations of the full ghost field and -vertex anomalous dimensions $\gammaCC$ and $\gammaCCG$ \cite{Luthe:2017ttc}, we have now also calculated the remaining $\Nf^{\{2,1,0\}}$ terms of the 5-loop gluon field renormalization constant $\gammaGG$ in Feynman gauge. 
In terms of the renormalized gauge coupling $a$ as defined in \eq\nr{eq:a}, we have obtained
\ba
\la{eq:g3}
\gammaGG &=& -a\,\Big[ 
\frac{8\nf-(10+3\xi)}{6}
+\gamma_{31} a
+\gamma_{32} a^2
+\gamma_{33} a^3
+\gamma_{34} a^4
+\dots
\Big] 
\;.
\ea
The coefficients $\gamma_{3n}$ are functions of the group invariants and the gauge parameter, see \app\ref{app:g3} for expressions up to four loops. At five loops and in Feynman gauge $\xi=0$, we have obtained (to clearly expose the group structure, we use a scalar-product-like notation, where e.g.\ $\{\cf^2,\cf,1\}.\{a,b,c\}=\cf^2 a+\cf b+c$)
\ba
2^{13} 3^5\,\gamma_{34} &=&\gamma_{344}\,\big[16\nf\big]^4 +\gamma_{343}\,\big[16\nf\big]^3 + \gamma_{342}\,\big[16\nf\big]^2 + \gamma_{341}\,\big[16\nf\big] + \gamma_{340} +\order{\xi} \;,\\
\gamma_{344} &=& \big\{\cf, 1\big\}.\big\{
107 + 144 \zeta_3, 
-619/2 + 432 \zeta_4
\big\}\;,\\
\gamma_{343} &=& \big\{\cf^2, \cf, d_1, 1\big\}.\big\{
576 (4961/48 -238 \zeta_3 + 99 \zeta_4), 
  576 (16973/288 + 221 \zeta_3 - 198 \zeta_4 + 72 \zeta_5), 
\nonumber\\&&
  -10368 (55/3 - 41 \zeta_3 + 12 \zeta_4 + 20 \zeta_5), 
  144 (14843/36 + 722 \zeta_3 + 165 \zeta_4 - 816 \zeta_5)
  \big\}\;,\\
\gamma_{342} &=& \big\{\cf^3, \cf^2, \cf d_1, \cf, d_2, d_1, 1\big\}.\big\{
82944 (2509/48 + 67 \zeta_3 - 145 \zeta_5), 
\nonumber\\&&
-1152 (135571/16 + 
    4225 \zeta_3 - 3024 \zeta_3^2 - 99 \zeta_4 - 18900 \zeta_5 + 
    5400 \zeta_6), 
\nonumber\\&&
    -6635520 (13/8 + 2 \zeta_3 - 5 \zeta_5), 
\nonumber\\&&
 288 (476417/72 - 23035 \zeta_3 - 25056 \zeta_3^2 + 34929 \zeta_4 - 
    44640 \zeta_5 + 10800 \zeta_6), 
\nonumber\\&&
 13824 (230 - 2354 \zeta_3 + 54 \zeta_3^2 + 360 \zeta_4 - 295 \zeta_5 + 
    225 \zeta_6), 
\nonumber\\&&
 6912 (2373 - 4715 \zeta_3 + 288 \zeta_3^2 + 900 \zeta_4 - 
    820 \zeta_5), 
\nonumber\\&&
    -72 (1524019/8 - 33931 \zeta_3 - 47808 \zeta_3^2 + 
    108225 \zeta_4 - 73572 \zeta_5 - 39600 \zeta_6)
            \big\}\;,\\
\gamma_{341} &=& \big\{\cf^4, \cf^3, \cf^2, \cf d_2, \cf, d_3, d_2, 1\big\}.\big\{
  20736 (4157 + 768 \zeta_3), 
\nonumber\\&&
  -165888 (11277/4 + 1541 \zeta_3 + 335 \zeta_5 - 2520 \zeta_7), 
\nonumber\\&&
 1152 (2208371/3 + 396403 \zeta_3 + 91800 \zeta_3^2 - 65115 \zeta_4 - 
    647460 \zeta_5 + 229500 \zeta_6 - 362880 \zeta_7), 
\nonumber\\&&
 165888 (236 - 386 \zeta_3 - 216 \zeta_3^2 - 895 \zeta_5 - 
    357 \zeta_7), 
\nonumber\\&&
    -5184 (1139437/9 - 29587 \zeta_3 + 18744 \zeta_3^2 + 
    42880 \zeta_4 - 124360 \zeta_5 + 25500 \zeta_6 - 
    33362 \zeta_7), 
\nonumber\\&&
    -1728 (11659/2 - 116251 \zeta_3 + 8880 \zeta_3^2 + 
    171 \zeta_4 + 59980 \zeta_5 + 40200 \zeta_6 - 
    99099 \zeta_7), 
\nonumber\\&&
    -1728 (77920 - 735952 \zeta_3 - 61272 \zeta_3^2 + 
    150480 \zeta_4 + 249580 \zeta_5 + 76500 \zeta_6 + 52479 \zeta_7), 
\nonumber\\&&
 72 (124662829/18 - 4899045 \zeta_3 - 63192 \zeta_3^2 + 3669873 \zeta_4 + 
    4836692 \zeta_5 - 2278200 \zeta_6 
\nonumber\\&&    
    - 4098024 \zeta_7)
     \big\}\;,\\
\gamma_{340} &=& \big\{d_3, 1\big\}.\big\{
6912 (47317 - 814000 \zeta_3 + 15294 \zeta_3^2 + 42300 \zeta_4 + 
    61390 \zeta_5 + 427125 \zeta_6 +\! 358848 \zeta_7), 
\nonumber\\&&    
    -144 (112182361/9 - 
    12985044 \zeta_3 - 2403444 \zeta_3^2 + 6431460 \zeta_4 + 
    53855480 \zeta_5 - 12870750 \zeta_6 
\nonumber\\&&        
    - 30266775 \zeta_7)
         \big\}\;.
\ea
From the first of \eq\nr{eq:gammaGG}, using the relation $\beta=2(\gammaCCG-\gammaCC)-\gammaGG$, this enables us to obtain the corresponding terms of the Beta function, whose coefficients we define as
\ba
\partial_{\ln\mu^2}\,a=-a\Big[\ep-\beta\Big]=-a\Big[\ep+b_0\,a+b_1\,a^2+b_2\,a^3+b_3\,a^4+b_4\,a^5+\dots\Big]
\;.
\ea
The $L$\/-loop coefficients $b_{L-1}$ are polynomials in $\nf$, and up to four loops read
\ba
3^1\,b_0 &=& \big[\!-4\big]\nf+11 \;,\\
3^2\,b_1 &=& \big[\!-36\cf-60\big]\nf+102\;,\\
3^3\,b_2&=& \big[132\cf+158\big]\nf^2+\big[54\cf^2-615\cf-1415\big]\nf+2857/2\;,\\
3^5\,b_3&=& \big[1232\cf+424\big]\nf^3
+432(132\zeta_3-5)d_3
+(150653/2-1188\zeta_3)
+\\\nonumber
&& \big[72(169-264\zeta_3)\cf^2+64(268+189\zeta_3)\cf
+1728(24\zeta_3-11)d_1
+6(3965+1008\zeta_3)
\big]\nf^2
+\\\nonumber
&& \big[11178\cf^3+\!36(264\zeta_3\!-\!1051)\cf^2+\!(7073\!-\!17712\zeta_3)\cf 
+\!3456(4\!-\!39\zeta_3)d_2
+\!3(3672\zeta_3\!-\!39143)
\big]\nf,
\ea
while at five loops we get (using the same scalar-product-like notation as above)
\ba
\la{eq:result}
3^5\,b_4 &=&b_{44}\,\nf^4+b_{43}\,\nf^3+b_{42}\,\nf^2+b_{41}\,\nf+b_{40} \;,\\
b_{44} &=& \Big\{\cf,1\Big\}.\Big\{-8(107+144\zeta_3),4(229-480\zeta_3)\Big\} \;,\\
b_{43} &=& \Big\{\cf^2,\cf,d_1,1\Big\}.\Big\{
-6(4961-11424\zeta_3+4752\zeta_4),
-48(46+1065\zeta_3-378\zeta_4),
\nonumber\\&&
1728(55-123\zeta_3+36\zeta_4+60\zeta_5), 
-3(6231+9736\zeta_3-3024\zeta_4-2880\zeta_5)
\Big\} \;,\\
\la{eq:b42}
b_{42} &=& \Big\{\cf^3, \cf^2, \cf d_1, \cf, d_2, d_1, 1\Big\}. \Big\{
-54 (2509 + 3216 \zeta_3 - 6960 \zeta_5), 
\nonumber\\&&
9(94749/2 - 28628 \zeta_3 + 10296 \zeta_4 - 39600 \zeta_5), 
25920 (13 + 16 \zeta_3 - 40 \zeta_5), 
\nonumber\\&&
3(5701/2 + 79356 \zeta_3 - 25488 \zeta_4 + 43200 \zeta_5), 
-864 (115 - 1255 \zeta_3 + 234 \zeta_4 + 40 \zeta_5), 
\nonumber\\&&
-432 (1347 - 2521 \zeta_3 + 396 \zeta_4 - 140 \zeta_5), 
843067/2 + 166014 \zeta_3 - 8424 \zeta_4 - 178200 \zeta_5\Big\} \;,\\
b_{41} &=& \Big\{\cf^4, \cf^3, \cf^2, \cf d_2, \cf, d_3, d_2, 1\Big\}. \Big\{
-81(4157/2 + 384 \zeta_3), 
81 (11151 + 5696 \zeta_3 - 7480 \zeta_5), 
\nonumber\\&&
-3 (548732 + 151743 \zeta_3 + 13068 \zeta_4 - 346140 \zeta_5), 
-25920 (3 - 4 \zeta_3 - 20 \zeta_5), 
\\&&
8141995/8 + 35478 \zeta_3 + 73062 \zeta_4 - 706320 \zeta_5, 
216 (113 - 2594 \zeta_3 + 396 \zeta_4 + 500 \zeta_5), 
\nonumber\\&&
216 (1414 - 15967 \zeta_3 + 2574 \zeta_4 + 8440 \zeta_5), 
-5048959/4 + 31515 \zeta_3 - 47223 \zeta_4 + 298890 \zeta_5\Big\} \;,
\nonumber\\
\la{eq:b40}
b_{40} &=& \Big\{d_3,1\Big\}. \Big\{-162 (257-9358 \zeta_3+1452 \zeta_4+7700 \zeta_5),
\nonumber\\&&
8296235/16 - 4890 \zeta_3 + 9801 \zeta_4/2 - 28215 \zeta_5\Big\} \;.
\ea
Out of these 5-loop coefficients, $b_{44}$ has in fact been known already for quite some time from a large-\/$\Nf$ analysis \cite{PalanquesMestre:1983zy,Gracey:1996he}, while $b_{43}$ was given in \cite{Luthe:2016ima}, as a proof-of-concept of our setup that we have used in this and earlier works \cite{Luthe:2016xec,Luthe:2017ttc}. 
The three coefficients $b_{42}$, $b_{41}$ and $b_{40}$ have first been computed by an independent group \cite{Herzog:2017ohr}, using the background field method, infrared rearrangement \cite{Vladimirov:1977ak} and the so-called $R^*$ operation \cite{Chetyrkin:1984xa} in order to map UV divergences onto the class of massless four-loop two-point functions which were evaluated via their code \code{FORCER} \cite{Ueda:2016sxw,Ueda:2016yjm,Ruijl:2017cxj}. Equations \nr{eq:b42}-\nr{eq:b40} fully coincide with the results of \cite{Herzog:2017ohr}.
As a further check of the 5-loop expressions given above, all coefficients reduce to the results given in \cite{Baikov:2016tgj} when setting the group invariants to their SU(3) values (cf.\ \eq\nr{eq:sun}).

To summarize, \eqs\nr{eq:result}-\nr{eq:b40} are in complete agreement with the corresponding terms of the Beta function given in \cite{Herzog:2017ohr}. This represents the first independent check of the correctness of this important renormalization group parameter. As a result, all terms of the five-loop Beta function have now been checked by two independent groups employing completely different methods, which should lead to a high confidence in its correctness.

%%%%%%%%%%%%%%%%%%%%%%%% SECTION %%%%%%%%%%%%%%%%%%%%%%
%
\section{Beyond the Feynman gauge}
\la{se:beyond}

To provide -- other than the confirmation of the five-loop Beta function presented in the previous section -- some genuinely new results in this paper, and to also showcase the versatility of our integral reduction codes \code{Crusher} and \code{Spades}, we have evaluated the linear terms of an expansion around the Feynman gauge ($\xi=0$). From the integral reduction point of view, this means that compared to the Feynman-gauge calculations, we need to be able to reduce integrals with one more dot and one more scalar product. 

We need to perform this exercise for a minimal set of three (out of five linearly independent) anomalous dimensions only, since $\beta$ and $\gamma_m$ are gauge parameter independent and already known \cite{Luthe:2016ima,Herzog:2017ohr,Luthe:2016xec}. The full set is then obtained via well-known linear relations, see \eqs\nr{eq:gammaGG}, \nr{eq:gammaGGGG}. 
For the convenience of the reader, we have prepared  computer-readable files that contain the complete set of renormalization constants and anomalous dimensions up to five loops including these new terms \cite{files}. 

In the following, we present our new five-loop results for these linear terms in $\xi$ for the ghost field, ghost-gluon vertex, as well as quark field anomalous dimensions. We keep the notation in line with our previous publications, such that it suffices to record the new terms here. 

%%%%%%%%%%%%%%%%%%%%%%%% SECTION %%%%%%%%%%%%%%%%%%%%%%
%
\subsection{Ghost field anomalous dimension}
\la{se:gammaCC}

As in \cite{Luthe:2017ttc}, we write the ghost field anomalous dimension as
\ba
\la{eq:123loop}
\gammaCC &=& -a \Big[ -\tfrac14(2 + \xi)
+ \gammaCc{1} a+ \gammaCc{2} a^2 + \gammaCc{3} a^3+ \gammaCc{4} a^4
+ \dots\Big] 
\;.
\ea
As fully gauge-dependent expressions up to four loops and the Feynman-gauge result at five loops have been given in \cite{Luthe:2017ttc}, we here add as a new result the linear term in $\xi$ at five loops:
\ba
\la{eq:gammaCc4}
2^{14}\,3^5\,\gammaCc{4} &=& \gammaCc{44}\,[16\nf]^4+\gammaCc{43}\,[16\nf]^3+\gammaCc{42}\,[16\nf]^2+\gammaCc{41}\,[16\nf]+\gammaCc{40} \;,\\
\gammaCc{4i} &=& \gammaCc{4i0} + \xi\,\gammaCc{4i1} + \order{\xi^2}\;,\\
\gammaCc{441} &=& 0\;,\\
\gammaCc{431} &=& \Big\{\cf, 1\Big\}.\Big\{0, 2(569 + 576 \zeta_3 - 1296 \zeta_4)\Big\}\;,\\
\gammaCc{421} &=& \Big\{\cf^2, \cf, d_1, d_2, 
  1\Big\}.\Big\{0, 36(-8191 + 6984 \zeta_3 + 1944 \zeta_4 - 3456 \zeta_5),0,0, 
  \nonumber\\&&
  -2 (66745 + 295182 \zeta_3 - 23328 \zeta_4 - 208764 \zeta_5) \Big\}\;,\\
\gammaCc{411} &=& \Big\{\cf^3, \cf^2, \cf d_2, \cf, d_2, d_3, 
  1\Big\}.\Big\{  0, -5184 (1349 + 3018 \zeta_3 - 720 \zeta_3^2 + 666 \zeta_4 - 2520 \zeta_5 -\! 1800 \zeta_6),
\nonumber\\&&
      0, 144 (90827 + 34092 \zeta_3 + 7776 \zeta_3^2 - 15552 \zeta_4 - 32832 \zeta_5 - 32400 \zeta_6), 
\nonumber\\&&
     5184 (32 + 4008 \zeta_3 + 432 \zeta_4 - 3060 \zeta_5 - 900 \zeta_6 - 1323 \zeta_7), 
\nonumber\\&&
     2592 (208 - 1141 \zeta_3 + 162 \zeta_3^2 - 297 \zeta_4 - 8375 \zeta_5 + 3525 \zeta_6 + 882 \zeta_7),
\\&&\nonumber
     4 (3979604 + 2404521 \zeta_3 - 750222 \zeta_3^2 + 1808649 \zeta_4 - 4632336 \zeta_5 - 1111725 \zeta_6 + 904932 \zeta_7) 
\!\Big\}\,,\\
\gammaCc{401} &=& \Big\{d_3, 
  1\Big\}.\Big\{ 10368 (2732 -\! 13091 \zeta_3 -\!  4146 \zeta_3^2 -\!  2241 \zeta_4 +\!  150485 \zeta_5 -\!  50925 \zeta_6 -\!  14434 \zeta_7),
\\&&\nonumber
-144(55138033/36 \!-\! 72901 \zeta_3 \!-\! 105498 \zeta_3^2 \!+\! 1074645 \zeta_4 \!-\! 1516578 \zeta_5 \!-\! 467775 \zeta_6 \!+\! 68397 \zeta_7)
\Big\}\;.
\ea
We observe that only 10 of the 17 possible color structures contain terms linear in $\xi$. 

%%%%%%%%%%%%%%%%%%%%%%%% SECTION %%%%%%%%%%%%%%%%%%%%%%
%
\subsection{Ghost-gluon vertex anomalous dimension}
\la{se:gammaCCG}

Again following the notation of \cite{Luthe:2017ttc}, the anomalous dimension of the ghost-gluon vertex is
\ba
\la{eq:gammaCCG}
\gammaCCG &=& - a (1 - \xi) \Big[
\tfrac12 
+ \tfrac{6-\xi}{8}\,a
+ \gammaCcg{2} a^2 + \gammaCcg{3} a^3 + \gammaCcg{4} a^4 + \dots \Big] \;,
\ea
where the prefactor follows from the finiteness of the Landau-gauge ghost vertex \cite{Taylor:1971ff,Blasi:1990xz}, which therefore does not need to be renormalized, hence $\gammaCCG|_{\xi=1}=0$.
The 3- and 4-loop coefficients (with full gauge dependence) and the 5-loop term (in Feynman gauge) have been given in \cite{Luthe:2017ttc}, to which we here add the linear term in $\xi$ at five loops:
\ba
\la{eq:gammaCcg4}
2^{14}\,3^5\,\gammaCcg{4} &=& 
\gammaCcg{43}\,[16\nf]^3+\gammaCcg{42}\,[16\nf]^2+\gammaCcg{41}\,[16\nf]+\gammaCcg{40} \;,\\
\gammaCcg{4i} &=& \gammaCcg{4i0} + \xi\,\gammaCcg{4i1} + \order{\xi^2}\;,\\
\gammaCcg{431} &=& 0 \;,\\
\gammaCcg{421} &=& \Big\{\cf, 1\Big\}.\Big\{ 0, 2(7855 - 22464 \zeta_3 + 3240 \zeta_4) \Big\}\;,\\
\gammaCcg{411} &=& \Big\{\cf^2, \cf, d_2, d_3, 1\Big\}.\Big\{ 0, 93312(56 - 31  \zeta_3 - 14  \zeta_4), 0, 
 \nonumber\\&&
5184(820 \zeta_3 - 78 \zeta_3^2 - 171 \zeta_4 - 660 \zeta_5 + 225 \zeta_6), 
 \nonumber\\&&
216(1247753/108 + 24604 \zeta_3 - 66 \zeta_3^2 + 1491 \zeta_4 - 8760 \zeta_5 + 1575 \zeta_6) \Big\}\;,\\
\la{eq:gammaCcg40}
\gammaCcg{401} &=& \Big\{d_3,1\Big\}.\Big\{ 5184(1986 - 74900 \zeta_3 + 4992 \zeta_3^2 + 10044 \zeta_4 + 52440 \zeta_5 - 24000 \zeta_6 + 
 25137 \zeta_7), 
 \nonumber\\&&
 -216(39394519/54 + 616864 \zeta_3 + 11472 \zeta_3^2 - 36984 \zeta_4 - 718836 \zeta_5 + 
 81300 \zeta_6 + 29925 \zeta_7) \!\Big\}\,.\nonumber
\ea

%%%%%%%%%%%%%%%%%%%%%%%% SECTION %%%%%%%%%%%%%%%%%%%%%%
%
\subsection{Quark field anomalous dimension}
\la{se:gammaQQ}

Following the notation of \cite{Luthe:2016xec}, the quark field anomalous dimension reads
\ba
\gammaQQ &=& 
-\cf\,a\,\Big[(1-\xi)+\gamma_{21}\,a+\gamma_{22}\,a^2+\gamma_{23}\,a^3+\gamma_{24}\,a^4+\dots\Big]\;.
\ea
As a new result, we add here the linear terms in the gauge parameter at five loops,
\ba
24^3\,\gamma_{24} &=& 
\gammaQq{44}\,\Big[16\nf\Big]^4
+\gammaQq{43}\,\Big[16\nf\Big]^3
+\gammaQq{42}\,\Big[16\nf\Big]^2
+\gammaQq{41}\,\Big[16\nf\Big]
+\gammaQq{40}\;,\\
\gammaQq{4i} &=& \gammaQq{4i0} + \xi\,\gammaQq{4i1} + \order{\xi^2}\;,\\
\gammaQq{441} &=& 0 \;,\\
\gammaQq{431} &=& \Big\{\cf, 1\Big\}.\Big\{ 0, 3197/144 + 7\zeta_3 - 36\zeta_4 \Big\}\;,\\
\gammaQq{421} &=& \Big\{\cf^2, \cf, d_1, 1\Big\}.\Big\{ 
0, 
-36 (23831/144 - 241 \zeta_3 + 54 \zeta_4 + 48 \zeta_5), 
0, 
\nonumber\\&&
-3541/2 - 13261 \zeta_3 + 3258 \zeta_4 + 4800 \zeta_5 
\Big\}\;,\\
\gammaQq{411} &=& \Big\{\cf^3, \cf^2, \cf d_1, \cf, d_1, d_2, 1\Big\}.\Big\{ 
0,
-432 (4261/12 + 947 \zeta_3 - 120 \zeta_3^2 + 111 \zeta_4 - 1140 \zeta_5 - 300 \zeta_6), 
0,
\nonumber\\&&
216 (1085843/432 - 1049 \zeta_3 - 324 \zeta_3^2 + 402 \zeta_4 - 924 \zeta_5 - 350 \zeta_6), 
576 (184 \zeta_3 + 168 \zeta_3^2 - 441 \zeta_7), 
\nonumber\\&&
48 (84 + 13124 \zeta_3 + 1596 \zeta_3^2 - 774 \zeta_4 - 10880 \zeta_5 + 1650 \zeta_6 - 9261 \zeta_7), 
\nonumber\\&&
4190641/12 + 652599 \zeta_3 + 40944 \zeta_3^2 - 106038 \zeta_4 - 390506 \zeta_5 - 1950 \zeta_6 + 5292 \zeta_7 
\Big\}\;,\\
\gammaQq{401} &=& \Big\{\cf^4, \cf^3, \cf^2, \cf d_2, \cf, d_2, d_3, 1\Big\}.\Big\{ 
0, 
0, 
864 (412 - 2480 \zeta_3 - 72 \zeta_3^2 + 1700 \zeta_5 + 1239 \zeta_7), 
\nonumber\\&&
1728 (120 - 1664 \zeta_3 + 1608 \zeta_3^2 + 2240 \zeta_5 - 2765 \zeta_7), 
\nonumber\\&&
-36 (20100 - 168816 \zeta_3 + 1944 \zeta_3^2 + 6804 \zeta_4 + 127120 \zeta_5 - 16200 \zeta_6 + 83657 \zeta_7), 
\nonumber\\&&
-24 (6960 + 286640 \zeta_3 + 138864 \zeta_3^2 - 145008 \zeta_4 - 534680 \zeta_5 + 334200 \zeta_6 - 355635 \zeta_7), 
\nonumber\\&&
 72 (5679 - 98586 \zeta_3 - 72432 \zeta_3^2 - 15714 \zeta_4 + 52640 \zeta_5 + 20100 \zeta_6 + 113666 \zeta_7), 
\nonumber\\&&
-4 (19063201/9 + 2617006 \zeta_3 + 164922 \zeta_3^2 - 624285 \zeta_4 - 3552844 \zeta_5 + 535650 \zeta_6 
\nonumber\\&&
+ 1691907 \zeta_7/4)
\Big\}\;.
\ea
As a check, the coefficients $\gammaQq{441}$ and $\gammaQq{431}$ coincide with the respective contribution extracted from the all-order large-\/$\Nf$ Landau gauge result \cite{Gracey:1993ua,Ciuchini:1999cv,Ciuchini:1999wy}.

%%%%%%%%%%%%%%%%%%%%%%%% SECTION %%%%%%%%%%%%%%%%%%%%%%
%
\section{Conclusions}
\la{se:conclu}

The purpose of this paper has been twofold. First, and perhaps most importantly, we have provided the first independent verification of the five-loop result for the Beta function of a non-abelian gauge field coupled to a single, but general, representation of a family of $\Nf$ fermions, first obtained in \cite{Herzog:2017ohr}. Finding full agreement, this constitutes an important check of this key result. As explained in \secs\ref{se:uv} and \ref{se:beta}, we have employed very different methods than those used in \cite{Herzog:2017ohr}, thus drastically increasing the confidence in the final result.

Second, we have used our setup to provide, for the complete set of 5-loop renormalization constants, the subleading terms in an expansion around the Feynman gauge. We have explicitly given the linear terms in the gauge parameter for a minimal choice of anomalous dimensions, while all others can be obtained from simple linear relations, see \eqs\nr{eq:gammaGG}-\nr{eq:gammaGGGG}; for completeness, we provide electronic versions of all renormalization constants in ancillary files \cite{files}.
This step aims at completing the five-loop renormalization program, and should be extended to include complete gauge dependence. As a crosscheck, it would be valuable to evaluate the linear gauge terms of another renormalization constant, for example the one of the gluon field. Due to the complexity of the 5-loop gluon propagator and limited computer resources to our disposal, we leave this as a future project.

It would be interesting to perform a 5-loop check of the known Landau-gauge relation \cite{Gracey:2002yt,Dudal:2002pq} between $\beta$, $\gammaGG$ and the anomalous dimension of the composite gauge-field operator $A_\mu^aA^{\mu a}$ (for results concerning renormalization of this operator up to 4 loops, see \cite{Gracey:2002yt,Dudal:2003np,Chetyrkin:2004mf,Chetyrkin:2009kh}). We have not yet attempted to do this (requiring full gauge dependence). In principle, this should be within reach of our method; having all master integrals at hand, it would in practice require substantial computer resources for an enlargement of our integral reduction tables, in order to accommodate integrals with larger exponents of propagators as well as numerators.

%%%%%%%%%%%%%%%%%%%%%%%% SECTION %%%%%%%%%%%%%%%%%%%%%%
%
\acknowledgments

The work of T.L.\ has been supported in part by DFG grants GRK 881 and SCHR 993/2; he is grateful to the theory group of the University of Bielefeld for hospitality, and for continued access to parts of their computer clusters, facilitated by the group of D.~B\"odeker.
A.M.\ is supported by a European Union COFUND/Durham Junior Research Fellowship under EU grant agreement number 267209.
P.M.\ was supported in part by the EU Network HIGGSTOOLS PITN-GA-2012-316704.
Y.S.\ acknowledges support from FONDECYT project 1151281 and UBB project GI-172309/C.

Note added: After submission of this work our results were confirmed independently in Ref.\ \cite{Chetyrkin:2017bjc}.

%%%%%%%%%%%%%%%%%%%%%%%% SECTION %%%%%%%%%%%%%%%%%%%%%%
%
\appendix
\section{Gauge field anomalous dimension}
\la{app:g3}

Let us give those coefficients of the gauge field anomalous dimension $\gammaGG$ of \eq\nr{eq:g3} that have not yet been listed in \se\ref{se:beta}. These lower-order terms have of course been known for a longer time already, and we refer to \cite{Luthe:2017ttc} for the corresponding references, as well as for explicit expressions in computer-readable form (where also the 5-loop terms in Feynman gauge had in fact already been included, assuming the validity of the Beta function \cite{Herzog:2017ohr}). 
From two to four loops, the coefficients of \eq\nr{eq:g3} read, with full gauge dependence:
\ba
2^4\,\gamma_{31} &=& 
\big(4 \cf + 5\big)\big[16\nf\big] 
-2\big(46 + 15 \xi - 2 \xi^2\big)
\;,\\
2^6 3^2\,\gamma_{32} &=& 
-\big(11 \cf + 19\big)\big[16\nf\big]^2
-2\big(36 \cf^2-(5 + 432 \zeta_3)\cf-(875 + 36 \xi - 324 \zeta_3)\big)\big[16\nf\big]
\nonumber\\&&
-2\big(8102 + 2286 \xi - 486 \xi^2 + 63 \xi^3 - 54 (4 - \xi) (2 - \xi) \zeta_3\big)
\;,\\
2^{9}3^5\,\gamma_{33} &=& \gamma_{333}\,\big[16\nf\big]^3 + \gamma_{332}\,\big[16\nf\big]^2 + \gamma_{331}\,\big[16\nf\big] + \gamma_{330} \;,\\
\gamma_{333} &=& \big\{\cf, 1\big\}.\big\{
-154, 
-355/2 + 216 \zeta_3
\big\} \;,\\
\gamma_{332} &=& \big\{\cf^2, \cf, d_1, 1\big\}.\big\{
-144 (169 - 264 \zeta_3), 
-4 (7541 + 15768 \zeta_3 - 5832 \zeta_4), 
\nonumber\\&&
3456 (11 - 24 \zeta_3), 
-41273 + 1229 \xi + 216 (85 - 6 \xi) \zeta_3 - 17496 \zeta_4
\big\} \;,\\
\gamma_{331} &=& \big\{\cf^3, \cf^2, \cf, d_2, 1\big\}.\big\{
-357696, 
144 (10847 + 5880 \zeta_3 - 12960 \zeta_5), 
\nonumber\\&&
-4 (363565 - 69903 \xi -6048 (89 - 9 \xi) \zeta_3 + 11664 (21 + \xi) \zeta_4 - 233280 \zeta_5), 
\nonumber\\&&
-6912 (64 - 516 \zeta_3 - 135 \zeta_5), 
  2809922 + 70690 \xi - 10449 \xi^2 
\nonumber\\&&
  - 648 (3855 - 296 \xi + 4 \xi^2) \zeta_3 
  + 972 (774 + 26 \xi + \xi^2) \zeta_4 + 855360 \zeta_5
\big\} \;,\\
\gamma_{330} &=& \big\{d_3, 1\big\}.\big\{
216 ( 4 (524 + 162 \xi - 27 \xi^2) -36(2456/3 + 556 \xi - 63 \xi^2 + 9 \xi^3 - 2 \xi^4)\zeta_3 
\nonumber\\&&
-45(2144 - 132 \xi^2 + 26 \xi^3 - \xi^4)\zeta_5), 
\nonumber\\&&
-2(8076320 + 3078806 \xi - 619326 \xi^2 + 134217 \xi^3 - 14580 \xi^4)
\nonumber\\&&
 + 162(66880 - 18672 \xi + 2534 \xi^2 - 36 \xi^3 - 37 \xi^4)\zeta_3
 -1944(444 - 292 \xi + 49 \xi^2 - 3 \xi^3)\zeta_4
\nonumber\\&&
 -405(32800 - 4224 \xi - 84 \xi^2 + 130 \xi^3 - 23 \xi^4)\zeta_5
\big\} \;.
\ea
Up to the linear terms in $\xi$, the 4-loop coefficient $\gamma_{33}$ has already been given in \cite{Czakon:2004bu}; we find full agreement for these terms, of course.
In the above, we have generalized that result to include full gauge parameter dependence.

%%%%%%%%%%%%%%%%%%%%%%%% SECTION %%%%%%%%%%%%%%%%%%%%%%
%
\section{Auxiliary mass renormalization}
\la{app:ZM}

Here, we list results for the auxiliary mass counterterm $Z_{\m^2}$, as introduced in \se\ref{se:uv}. Let us stress once more that this term, being an artefact of our infrared regularization method, is non-physical (and hence also gauge-parameter dependent). Nevertheless, to render this paper self-contained, and in order to facilitate comparisons in future works, we wish to record its detailed form. In terms of the normalized gauge coupling $a$ of \eq\nr{eq:a}, its structure is
\ba
\la{eq:ZM}
Z_{\m^2} &=& \frac{a}{\ep}\Big(
\frac{[16\nf] + (4-3\xi)}{16} 
+ \frac{a}{\ep}\Big[ [16\nf]z_{11} + z_{10} \Big] 
+ \frac{a^2}{\ep^2}\Big[ [16\nf]^2z_{22} +[16\nf]z_{21} + z_{20} \Big] \nonumber\\&&
+ \frac{a^3}{\ep^3}\Big[ [16\nf]^3z_{33} +[16\nf]^2z_{32} +[16\nf]z_{31} + z_{30} \Big]  \nonumber\\&&
+ \frac{a^4}{\ep^4}\Big[ [16\nf]^4z_{44} +[16\nf]^3z_{43} +[16\nf]^2z_{42} +[16\nf]z_{41} + z_{40} \Big] 
+\dots\Big)\;,
\ea
where we have explicitly shown $\nf$\/-dependence, and where the coefficients $z_{ij}$ are degree-$i$ polynomials in $\ep$ that can be extracted from the $(i\!+\!1)$\/-loop gluon propagator at zero external momentum. In turn, the coefficients of the $z_{ij}$ then depend on the group invariants of \se\ref{se:color}, as well as the gauge parameter $\xi$.
Note that, due to the absence of a tree-level contribution, there is no leading constant as in all other RCs (which have the generic form $Z=1+az_1+a^2z_2+\dots$). After accounting for the different normalization conventions, the leading (1-loop) term of \eq\nr{eq:ZM} can be seen to coincide with the one given in \eq(4) of \cite{Chetyrkin:1997fm}. At two loops, again using the scalar-product-like notation of the main text to emphasize group structure (and normalizing conveniently with an overall factor), we have
\ba
2^{10}\, z_{11} &=& 
+\ep^1\,\big\{\cf,1\big\}.\big\{-4(1-\xi),5(5/3-\xi)\big\} 
\nonumber\\&&
+\ep^0\,\big\{\cf,1\big\}.\big\{8(1-\xi),-2(13-5\xi)\big\}
\;,\\ 
2^{10}\, z_{10} &=&
+\ep^1\,\big(2(146/3-49\xi+10\xi^2)\big)
\nonumber\\&&
+\ep^0\,\big(-2(72-58\xi+15\xi^2)\big)
\;.
\ea
The three-loop coefficients read
\ba
2^{15} 3\, z_{22} &=& 
+\ep^2\,\big\{\cf,1\big\}.\big\{20, -394/27\big\} 
\nonumber\\&&
+\ep^1\,\big\{\cf,1\big\}.\big\{-16, 160/9\big\} 
\nonumber\\&&
+\ep^0\,\big\{\cf,1\big\}.\big\{0,-40/3\big\} 
\;,\\ 
2^{15} 3\, z_{21} &=& 
+\ep^2\,\big\{\cf^2,\cf,1\big\}.\big\{-16 (2 - 2 \xi + \xi^2), -8 (253 - 79 \xi + 17 \xi^2 - 48(5 - \xi) \zeta_3),
\nonumber\\&&
    22768/27 - 184 \xi + 71 \xi^2 -48 (20 + \xi) \zeta_3\big\} 
\nonumber\\&&
+\ep^1\,\big\{\cf^2,\cf,1\big\}.\big\{-32 (7 - 2 \xi + \xi^2), 
    8 (186 - 93 \xi + 19 \xi^2), -2 (5534/9 - 399 \xi + 
       86 \xi^2)\big\} 
\nonumber\\&&
+\ep^0\,\big\{\cf^2,\cf,1\big\}.\big\{64 (1 - 2 \xi + \xi^2), -48 (10 - 13 \xi + 3 \xi^2), 
    4 (946/3 - 169 \xi + 35 \xi^2)\big\} 
\;,\\ 
2^{15} 3\, z_{20} &=&
+\ep^2\,\big( 8 (22837/27 - 837 \xi + 
       292 \xi^2 - 49 \xi^3 + 6 (12 - 22 \xi + 5 \xi^2) \zeta_3) \big)
\nonumber\\&&
+\ep^1\,\big( -4 (29168/9 - 2976 \xi + 1095 \xi^2 - 
       154 \xi^3) \big)
\nonumber\\&&
+\ep^0\,\big( 12 (6872/9 - 604 \xi + 222 \xi^2 - 
       35 \xi^3) \big)
\;.
\ea
At four loops, we get 
\ba
2^{21} 3^4\, z_{33} &=& 
+\ep^3\,\big\{\cf,1\big\}.\big\{-872, -1334/3 + 576 \zeta_3\big\} 
\nonumber\\&&
+\ep^2\,\big\{\cf,1\big\}.\big\{720, 204\big\} 
\nonumber\\&&
+\ep^1\,\big\{\cf,1\big\}.\big\{-288, 360\big\} 
\nonumber\\&&
+\ep^0\big\{\cf,1\big\}.\big\{0,-288\big\} 
\;,\\ 
2^{21} 3^4\, z_{32} &=& 
   +\ep^3\,\big\{\cf^2,\cf,1\big\}.\big\{216 (151 - 19 \xi - 128 \zeta_3), 
   18 (62026/9 - 1507 \xi - 96 (46 - 7 \xi) \zeta_3 
\nonumber\\&&
   + 
      288 (13 - \xi) \zeta_4), 
      -2 (9157/3 - 2957 \xi - 
      108 (219 - 31 \xi) \zeta_3 + 324 (65 + \xi) \zeta_4)\big\} 
\nonumber\\&&
+\ep^2\,\big\{\cf^2,\cf,1\big\}.\big\{576 (19 - 6 \xi), -108 (2723/3 - 239 \xi + 32 (11 + \xi) \zeta_3), 
   3 (17429 - 5443 \xi 
\nonumber\\&&
   + 144 (79 - \xi) \zeta_3)\big\} 
\nonumber\\&&
+\ep^1\,\big\{\cf^2,\cf,1\big\}.\big\{-864 (9 - 5 \xi), 72 (578 - 211 \xi), -36 (1881 - 368 \xi)\big\} 
\nonumber\\&&
+\ep^0\,\big\{\cf^2,\cf,1\big\}.\big\{0, -4752(1 - \xi), 
   36 (979 - 147 \xi) \big\} 
\;,\\ 
2^{21} 3^4\, z_{31} &=& 
+\ep^3\,\big\{\cf^3,\cf^2,\cf,d_2,1\big\}.\big\{864 (240 + 90 \xi - 3 \xi^2 + \xi^3 + 
      96 (1 - \xi) \zeta_3), 
\nonumber\\&&
   216 (8228 + 
   \!5 \xi (374 - 50 \xi + \xi^2) - 
      \!32 (187 + 42 \xi + 9 \xi^2) \zeta_3 - \!288 \xi (5 - \xi) \zeta_4 - 
      \!1920 (5 - 2 \xi) \zeta_5), 
\nonumber\\&&
      -18 (3699988/9 - 126068 \xi + 24669 \xi^2 - 4239 \xi^3 - 
      48 (13546 - 2347 \xi + 150 \xi^2) \zeta_3 
\nonumber\\&&
      + 
      144 (946 - 151 \xi + 6 \xi^2) \zeta_4 + 
      2880 (52 - 2 \xi + \xi^2) \zeta_5), 
\nonumber\\&&
      -5184 (2 (48 - 8 \xi + 
         \xi^2) - (1016 - 60 \xi - 28 \xi^2 + 11 \xi^3) \zeta_3 + 
      20 (74 - 21 \xi + 4 \xi^2) \zeta_5), 
\nonumber\\&&
   2587604/3 - 474148 \xi + 146205 \xi^2 - 60993 \xi^3 / 2 - 
      432 (10689 - 397 \xi - 60 \xi^2 + 4 \xi^3) \zeta_3 
\nonumber\\&&
      + 
      648 (2102 + 226 \xi - 7 \xi^2) \zeta_4 + 
      2160 (22 + \xi) (34 + 5 \xi) \zeta_5)\big\} 
\nonumber\\&&
+\ep^2\,\big\{\cf^3,\cf^2,\cf,d_2,1\big\}.\big\{1728 (43 - 5 \xi - 3 \xi^2 + 
      \xi^3), 
\nonumber\\&&
      -432 (12242/3 - 773 \xi + 186 \xi^2 - 25 \xi^3 - 
      96 (16 - 5 \xi + \xi^2) \zeta_3), 
\nonumber\\&&
   108 (189176/3 - 24678 \xi + 5943 \xi^2 - 839 \xi^3 - 
      16 (1906 - 751 \xi + 66 \xi^2) \zeta_3), 0, 
\nonumber\\&&
      -3 (706004 - 362512 \xi + 
      120456 \xi^2 - 20997 \xi^3 - 
      144 (3350 - 710 \xi - 67 \xi^2) \zeta_3)\big\} 
\nonumber\\&&
+\ep^1\,\big\{\cf^3,\cf^2,\cf,d_2,1\big\}.\big\{-3456 (1 - \xi) (16 - 
      2 \xi + \xi^2), 
   864 (752 - 642 \xi + 214 \xi^2 - 31 \xi^3), 
\nonumber\\&&
   -72 (36660 - 25388 \xi + 
      7443 \xi^2 - 969 \xi^3), 0, 
   54 (37136 - 25888 \xi + 9072 \xi^2 - 1289 \xi^3)\big\} 
\nonumber\\&&
+\ep^0\,\big\{\cf^3,\cf^2,\cf,d_2,1\big\}.\big\{6912 (1 - \xi)^3,
    -5184 (1 - \xi)^2 (18 - 5 \xi), 
\nonumber\\&&
    432 (1 - \xi) (1248 - 538 \xi + 
      89 \xi^2), 0, 
      -36 (33052 - 20088 \xi + 6678 \xi^2 - 
      945 \xi^3)\big\} 
\;,\\ 
2^{21} 3^4\, z_{30} &=&
      +\ep^3\,\big\{d_3,1\big\}.\big\{-1296 (12 (28 - 18 \xi + 5 \xi^2) - 
      4 (3128 - 1868 \xi + 329 \xi^2 - 69 \xi^3 + 18 \xi^4) \zeta_3 
\nonumber\\&&
      + 
      5 (2144 + 56 \xi - 716 \xi^2 + 182 \xi^3 - 9 \xi^4) \zeta_5), 
\nonumber\\&&
   2 (2 (11725124/3 - 3353810 \xi + 1154358 \xi^2 - 332937 \xi^3 + 43740 \xi^4) 
\nonumber\\&&
   + 
      18 (215728 - 195576 \xi + 38430 \xi^2 - 999 \xi^4) \zeta_3 - 
      648 (640 - 1084 \xi + 257 \xi^2 - 21 \xi^3) \zeta_4 
\nonumber\\&&
      - 
      135 (34240 - 13688 \xi - 1012 \xi^2 + 1078 \xi^3 - 
         207 \xi^4) \zeta_5)\big\} 
\nonumber\\&&
+\ep^2\,\big\{d_3,1\big\}.\big\{0,-12 (2045504 - 1866458 \xi + 
      795753 \xi^2 - 200016 \xi^3 + 23004 \xi^4 
\nonumber\\&&
      + 
      72 (1072 - 2380 \xi + 1049 \xi^2 - 
         147 \xi^3) \zeta_3)\big\} 
\nonumber\\&&
+\ep^1\,\big\{d_3,1\big\}.\big\{0,72 (348728 - 298948 \xi + 129576 \xi^2 - 
      31335 \xi^3 + 3294 \xi^4)\big\} 
\nonumber\\&&
+\ep^0\,\big\{d_3,1\big\}.\big\{0,-36 (318160 - 243624 \xi + 
      101484 \xi^2 - 25020 \xi^3 + 2835 \xi^4)\big\} 
\;.
\ea
From the Feynman-gauge gluon propagator that we have evaluated in order to compute the gluon field anomalous dimension $\gammaGG$ in \se\ref{se:beta}, we could in principle even extract the 5-loop terms $z_{4i}$ at $\xi=0$.
Since they do not enter the calculations presented in this paper, however, we do not list them here.

%%%%%%%%%%%%%%%%%%%%%%%% BIBLIO %%%%%%%%%%%%%%%%%%%%%%%
%

\end{document}